\documentclass[conference]{IEEEtran}
\IEEEoverridecommandlockouts
\usepackage{cite}
\usepackage{amsmath,amssymb,amsfonts}
\usepackage{algorithmic}
\usepackage{graphicx}
\usepackage{textcomp}
\usepackage{xcolor}
\begin{document}

\title{Feature Mining for Encrypted Malicious Traffic Detection with Deep Learning and Other Machine Learning Algorithms \\}

\author{\IEEEauthorblockN{Zihao Wang, Vrizlynn L. L. Thing}
\IEEEauthorblockA{\textit{Cybersecurity Strategic Technology Centre} \\
\textit{Singapore Technologies Engineering}\\
zihao.wang@stengg.com, vriz@ieee.org}
}
\maketitle
 
\begin{abstract}
The popularity of encryption mechanisms poses a great challenge to malicious traffic detection. The reason is traditional detection techniques cannot work without the decryption of encrypted traffic. Currently, research on encrypted malicious traffic detection without decryption has focused on feature extraction and the choice of machine learning or deep learning algorithms. In this paper, we first provide an in-depth analysis of traffic features and compare different state-of-the-art traffic feature creation approaches, while proposing a novel concept for encrypted traffic feature which is specifically designed for encrypted malicious traffic analysis. In addition, we propose a framework for encrypted malicious traffic detection. The framework is a two-layer detection framework which consists of both deep learning and traditional machine learning algorithms. Through comparative experiments, it outperforms classical deep learning and traditional machine learning algorithms, such as ResNet and Random Forest. Moreover, to provide sufficient training data for the deep learning model, we also curate a dataset composed entirely of public datasets. The composed dataset is more comprehensive than using any public dataset alone. Lastly, we discuss the future directions of this research.

\end{abstract}


\section{\bf Introduction}
With the widespread application of encryption technology and the increasing requirements for privacy and data security, more and more enterprises choose to use encryption mechanisms to hide their payload context, which has led to the explosion of encrypted traffic. Network traffic is the amount of data moving across a computer network at any given time, so it can also be referred to as data traffic that is broken down into packets and sent across the network and then reassembled by the receiving devices. As traffic is not encrypted, the data within the traffic is in plain text; as some encryption protocols are applied, the data within the traffic is encrypted and such traffic is called encrypted traffic. In our research, we focus on network traffic generated in the event of malicious activities, such as Trojan and Botnet, that have been encrypted using certain encryption techniques. According to Google's transparency report, the number of websites using encrypted traffic has grown from 50\% in 2014 to 95\% in 2022. 97\% of the world's top 100 websites use hypertext transfer protocol secure (HTTPS) protocols in 2022 [1]. While (data) encryption technology protects user privacy, the abuse of encryption technology is also profoundly changing the threat landscape of network security.  The abuse of encryption technology has not only made it easier for online fraud and illegal online transactions, but also for hackers to evade the detection of ransomware, phishing, and data breaches. According to a study by WatchGuard Technologies [2], in the second quarter of 2021, 91.5\% of malware detection involved malware arriving over HTTPS encrypted connections. This means that organizations that do not have a detection system to decrypt and scan HTTPS traffic for malware would miss nine-tenths of malware. Another study, Zscaler, also reported HTTPS threat to grow over 314\% in 2021, and the growth rate exceeded 250\% for the second straight year [3].

Traditional detection techniques (i.e., payload-based deep packet inspection (DPI) methods and port-based identification methods.) are often powerless in the face of malicious encrypted traffic. In this context, developing detection and defense technologies based on encrypted traffic is imperative. The methods of encrypted traffic detection can be divided into two categories. The first one is detecting traffic by obtaining plain text through decryption. The second one is detecting and identifying encrypted traffic based on non-decryption methods such as machine learning methods. However, since one of the reasons why encrypted traffic is widely used is the protection of user privacy, it is not recommended to detect encrypted traffic through decryption. Therefore, the current mainstream research direction is to detect encrypted malicious traffic without decryption based on different machine learning algorithms.

The detection and identification of encrypted traffic based on non-decryption methods mainly rely on machine learning technology. Benefiting from the rapid development of computer hardware in recent years, artificial intelligence has been applied in various fields and the performance is worth relying on. Machine learning is a branch of artificial intelligence (AI) and computer science which a computer uses algorithms to learn from data to complete a prediction or classification task without being explicitly programmed. Deep learning is a specialized subset of machine learning. It layers algorithms and computational units or neurons to implement artificial neural networks. Deep learning uses complex algorithmic structures modeled on the human brain. This makes it possible to process unstructured data, such as images, natural language processing, and sentiment analysis. Currently, research on malicious encrypted traffic detection mainly focuses on feature extraction and the selection of machine learning or deep learning algorithms.

In this paper, we first conduct an in-depth analysis of network traffic features and propose a new granularity for encrypted traffic features. We also propose an encrypted malicious traffic detection framework that is constructed by both traditional machine learning and deep learning algorithms. Furthermore, since deep learning models always require a large amount of training data, we also curate a large-size deep learning model training dataset. The newly composed dataset is more comprehensive than any single public dataset. Finally, current challenges and future directions are discussed.

The paper structure is organized as follows: we review the existing related works and techniques in Section 2. In section 3, we conduct a comprehensive analysis and classification of network traffic features. Meanwhile, we also propose our new feature creation approach specifically designed for encrypted traffic analysis and encrypted malicious traffic identification and classification. At the same time, we design a detection framework that is constructed by both deep learning and machine learning in Section 4. A series of comparative experiments are conducted and their performance evaluations are discussed and analyzed. Finally, we conclude the paper in Section 5, by discussing the remaining challenges and future directions.

\section{\bf Literature Review}
In recent years, the rise of artificial intelligence has allowed us to use machine learning and deep learning methods to detect encrypted malicious traffic without decryption. Previous experiments have proved that the detection results are accurate. Traffic Feature extraction and machine learning algorithms selection have become the main focuses in the research of encrypted malicious traffic detection.

Bazuhair et al. [4] proposed an encoding method for converting selected features of Transport Layer Security (TLS)/Secure Sockets Layer (SSL) traffic into 2D images, and data argumentation is performed through Perlin noise. The final generated image is used to train the convolutional Neural Network (CNN) binary detection model. 0.40\% false negative rate and 5.60\% false positive rate are achieved under the CTU-13 dataset of stratosphere Lab. TLS encrypted malicious traffic classification method based on Support Vector Machines (SVM) and CNN is proposed by Lucia and Cotton. [5]. 99.91\% accuracy and F1 are achieved by 1D-CNN with Adam optimizer and 99.97\% accuracy and F1 are achieved by SVM with radial basis function kernel. 

Zhang et al. [6] proposed a transfer learning method based on Efficientnet to detect encrypted malicious traffic. The author first pre-trained an Efficientnet model, Efficientnet-B0, through the imagenet dataset, and then transfer it to a small amount of encrypted traffic dataset for training. In this process, manual feature extraction by experts is omitted. The model also achieved 100\% detection accuracy and recall rate using only a small number of datasets. However, since the authors used small-size traffic datasets, the robustness of the model remains to be determined.

Lopez-Martin et al. [7] designed an IoT traffic classification model by combining both recurrent neural network (RNN) and CNN no matter whether the traffic is encrypted or non-encrypted.  The author provided a complete study on a series of CNN+RNN architectures, feature selection, and the length of traffic packets input. The classification model achieved 96.32\% accuracy with an imbalanced dataset. 5 feature sets are selected to analyze the importance of the features in model training. The threshold of packet number of each flow session is also considered to balance the computing time and classification performance. Through comparative experiments, packet numbers between five and fifteen outperform other selected packet numbers and 94.50\% accuracy and F1 are achieved. The author also selected to use the zero-padding method to ensure each session has the same number of packets if sessions with less than the pre-decided packets number. However, the zero-padding method is not reasonable for certain traffic features like inter-arrival time, windows length of each packet, and time to live of each packet, which may bring unpredictable bias to the model training.

Bader et al. [27] proposed a novel design (MalDIST) of the extension of the DISTILLER model [28] to the field of encrypted malicious traffic detection and classification. The authors provide an encrypted malware traffic detection and classification framework consisting of multiple deep-learning models, including 1D CNN, 2D CNN, and bidirectional GRU models. A novel modality by grouping traffic and calculating statistical features based on different groups is also provided for malicious traffic. Finally, the authors compare their proposed model with seven different algorithms. Their proposed MalDIST achieved 99.7\% Accuracy, precision, recall, and F1 which outperforms others.

Yao et al. [8] proposed two methods to classify encrypted traffic, using Long short-term memory (LSTM) with an attention mechanism or based on a hierarchical attention network (HAN). The author also conducted comparative experiments on traditional machine learning models such as decision trees, XGBoost, and deep learning algorithms such as 1D-CNN. After training with the VPN-non-VPN dataset, the author proposed their two detection models outperform decision tree models [16] as well as 1D CNN models [17].

Ferriyan et al. [9] proposed an encrypted malicious traffic detection (TLS2Vec) based on the TLS handshake and payload features. The advantage of their detection model is that there is no need to wait for the traffic session to finish, which ensures privacy to a certain extent. The authors generate words from selected features and train with LSTM and BiLSTM models. Malicious and benign traffic data from the CTU Malware facility Project [19] is selected and TLS2Vec performs better than Non-TLS2Vec, which uses neither symbols nor Word2Vec embedding. 

Bovenzi et al. [10] proposed a two-stage intrusion detection architecture for both known and unknown attacks, which is named H2ID. A novel multi-modal deep auto-encoder is designed to achieve lightweight anomaly detection. Then the detected traffic is classified into different attack traffic types based on soft-output classifiers. 

A novel cross-layer feature representation method under TLS and Internet Protocol Security (IPSec) protocols is proposed by Meghdouri et al. [11] in their Random Forest (RF) classification model. Three different public datasets are selected to test the model separately and achieved 100\%, 92.60\%, and 92\% F1 scores. Comparative experiments with other existing methods are also conducted under the same datasets. Their RF classification model outperforms other methods through several comparative experiments under the same datasets.

7 different machine learning algorithms are selected from Stergiopoulos et al. [12] to test the performance of their proposed Transmission Control Protocol (TCP) side channel features. 99.80\% accuracy under the decision tree method is achieved under a composed dataset from CTU-13 [18], FIRST [20], and Milicenso [21]. 99.80\% accuracy is achieved by the XGBoost model with machine-selected features in the research of Shekhawat et al. [13]

In [14], the authors proposed a distance-based framework for encrypted malicious traffic classification. The framework consists of a series of detection models for different malware. Gaussian mixture model (GMM) and ordering points to identify the clustering structure (OPTICS) are applied to calculate the distance between malware so as to determine the new malware class. Then, 24 XGBoost models are trained to classify 24 kinds of malware.

\section{\bf Traffic Feature Analysis}

In this Section, we further explored the hidden attributes of encrypted traffic. We also increased the dimension and quantity of encrypted traffic features so as to bring incremental information to the encrypted malicious traffic detection task. However, due to the lack of recognized feature standard definition and extraction logic in the field of network traffic feature analysis, we will first conduct a comprehensive analysis of network traffic features in this section.

According to the point of view in our previous article [22], traffic features can be divided into two main categories: protocol-agnostic numerical features and protocol-specific features. The protocol-specific features are features extracted for certain specific encryption protocols. These features are created from specific attributes of the specific protocol. Therefore, such features cannot be extracted from other protocols. For example, TLS/SSL version types, mean of public certificate key, and mean of certificate validity can be extracted in TLS/SSL protocol. However, these features cannot be extracted from traffic under the SSH protocol. Moreover, extracting protocol-specific features requires the use of some extraction tools such as ZeekIDS (BroIDS). Thus, the entire feature extraction process is time-consuming. For protocol-specific features, its main limitation is that such features are unique to some specific encryption protocols. Once other encryption protocol traffic is also included in the dataset, these features cannot be extracted from other encryption protocol traffic. 

The protocol-agnostic numerical features are features that are not limited by the encrypted protocol type of traffic. Such features can be extracted from both encrypted and unencrypted traffic as well, such as payload size, time to live, and flow duration. Furthermore, protocol-agnostic numerical features can be further divided into two granularities: packet-based features and session-based features. Packet-based features are extracted at the level of traffic packets, such as the payload length and inter-arrival time of forward/backward packets of each session. Session-based features are extracted at the flow session level. Such session-based features can only be extracted as one value in each flow session, such as the flow duration of each session and the total bytes of each session. In general, Protocol-agnostic features are more robust and easier to extract than protocol-specific features. However, due to the huge variety of such features, the extraction process often requires prior knowledge and consumes a lot of time.

Due to the encryption mechanism, the payload context can no longer be used as a signature to identify encrypted malicious traffic. The use of certain encryption protocol-specific features is also not enough to completely detect encrypted malicious traffic under different encryption protocols. The current research tends to use side-channel protocol-agnostic features as the feature set to train AI detection models. These features can be extracted no matter the traffic under which protocols. However, these features do not particularly represent the characteristics of encrypted traffic either because such features can be extracted whether the traffic is encrypted or not. Therefore, for encrypted traffic, the dimension of available features is decreasing significantly. Highly discriminatory features are even rarer. Therefore, one bottleneck in maintaining and improving the performance of encrypted malicious traffic detection is the insufficient meaningful feature number and feature mining. 

Adi [15] applied a series of state-of-the-art deep learning and machine learning algorithms to compare performance against malicious traffic detection. Their comparative experiments proved that in the field of malicious traffic detection, the performance of the machine learning algorithm is not necessarily worse than that of deep learning algorithms. They may get very similar detection performance or even better detection result in certain evaluation measures. For example, RF and M1 CNN models have obtained similar detection results in the comparative experiment of binary malicious traffic detection, and RF is even better than CNN models and other deep learning models based on certain evaluation measures such as accuracy, F1-score, and Recall. The authors' experiments discussed that in some cases, simpler and better-performing solutions with appropriate feature sets may exist. Therefore, it is worthwhile to carry out more in-depth feature mining on encrypted traffic rather than simply combining features that can be extracted regardless of whether the traffic is encrypted or not.


\paragraph{\bf \textit{Novel encrypted traffic feature creation approach}}

We propose a new concept of traffic feature creation approach by analyzing the peculiarities of encrypted traffic sessions and packets which is named the specific encrypted traffic feature (Enc Feature). It can be viewed as a new granularity of the encrypted network traffic feature. The Enc Feature is different from the widely used protocol-agnostic numerical feature (protocol-agnostic numerical features are extracted independently of whether the traffic is encrypted or not). It is to analyze encrypted traffic sessions and the encrypted packets in each such encrypted session to extract the session-based and packet-based Enc Features that only encrypted traffic will have.

Before the Enc Feature concept, the features chosen by researchers were the same protocol-agnostic numerical features for both non-encrypted and encrypted traffic analysis. Its extraction logic allows us to extract the features without considering whether the packets in the encrypted session are encrypted or not. This feature creation and selection logic actually reduces the importance of the feature itself in encrypted traffic analysis but focuses more on whether the original data itself has been encrypted or not. It is the encrypted traffic detection feature set if traffic in the dataset is encrypted, and can be viewed as the unencrypted traffic detection feature set if the traffic in the dataset is not encrypted. If viewed from the perspective of the traffic packet level, the features of both encrypted and unencrypted packets are extracted and mixed together at the same time. This also means that it cannot perfectly interpret the specific characteristics of encrypted traffic. Our proposed concept is the only work that is protocol-agnostic and at the same time considers both session level and packet level encrypted traffic features which we will elaborate more in the next sub-sections.


\paragraph{\bf \textit{Novel encrypted traffic feature extraction process}}

Enc Feature can be considered as the exclusive feature of encrypted traffic analysis, detection, and classification. The extraction logic for Enc Feature considers only encrypted traffic packets within an encrypted session and excludes non-encrypted packets within the encrypted session. We find that existing works proposing encrypted traffic analysis did not consider fully eliminating non-encrypted traffic packets. For example, encrypted traffic sessions belonging to protocols such as TLS contain non-encrypted handshake packets. Our work includes filtering off such non-encrypted traffic in order to extract features that represent purely encrypted traffic. The following two-step process describes our filtering methodology:

1. Filter out all non-encrypted traffic sessions in the mixed traffic dataset. (this ensures that there are no non-encrypted traffic sessions at the session level that will affect the encrypted traffic analysis.)

2. Filter out all non-encrypted traffic packets from the encrypted sessions. (make sure no non-encrypted attributes appear at both the session level and packet level.)

A total of 78 Enc features can be extracted from the processed traffic. The Enc Feature list is provided in Table I. Such features can be further classified into two groups: session level encrypted traffic features (No.4-23 features in Table I) and packet level encrypted traffic features (No.1-3,24 features in Table I). For each above groups, feature engineering is further applied to generate the min, max, mean, median, standard deviation, and variance values of each encrypted traffic feature (No. 25-78).

In addition, traditional protocol-agnostic numerical features still have considerable analytical value. Our proposed Enc features can be further analyzed in comparison with traditional protocol-agnostic numerical features to create ratio traffic features (No.79-143). Such feature engineering will generate a new traffic feature granularity to help researchers further analyze the distribution characteristics of encrypted traffic. We take flow duration as an example. Flow\_duration is a commonly used traditional protocol-agnostic numerical feature in existing work that is defined as the time difference between the last packet and the first packet of each session. The enc\_flow\_duration feature belongs to our proposed Enc Feature which is defined as the time difference between the last encrypted packet and the first encrypted packet in each session. The ratio between flow\_duration and enc\_flow\_duration can be calculated and applied to the model training. Ratio\_of\_IP\_packet\_length\ is the ratio between total\_length\_of\_IP\_packet and total\_length\_of\_enc\_IP\_packet.

\begin{table*}[h!]
\centering
\caption{Enc Features (specific encrypted traffic feature) List}
\begin{tabular}{|c|l|c|c|}
\hline
No.    & Enc Features                                                             & \begin{tabular}[c]{@{}l@{}}Time \\ based\end{tabular} & \begin{tabular}[c]{@{}l@{}}Statis \\ based\end{tabular} \\ \hline
1      & inter arrival time of forward enc traffic                                &  \checkmark           &                   \\ \hline
2      & inter arrival time of backward enc traffic                               &   \checkmark          &                   \\ \hline
3      & ratio to previous enc packet                                             &            &    \checkmark                \\ \hline
4      & flow duration enc                                                        &   \checkmark          &                   \\ \hline
5      & flow duration of backward enc traffic                                    &   \checkmark          &                   \\ \hline
6      & flow duration of forward enc traffic                                     &   \checkmark          &                   \\ \hline
7      & total time to live of forward enc traffic                                &   \checkmark          &                   \\ \hline
8      & total time to live of backward enc traffic                               &    \checkmark         &                   \\ \hline
9      & total TCP windows size value of forward enc traffic                      &            &   \checkmark                 \\ \hline
10     & total TCP windows size value of backward enc traffic                     &            &  \checkmark                  \\ \hline
11     & Total length of enc IP packet                                            &            &   \checkmark                 \\ \hline
12     & Total time to live of enc traffic                                        &       \checkmark      &                   \\ \hline
13     & Total length of forward enc payload                                      &            &     \checkmark               \\ \hline
14     & Total length of backward enc payload                                     &            &    \checkmark                \\ \hline
15     & Total length of forward enc IP header                                    &            &   \checkmark                 \\ \hline
16     & Total length of backward enc IP header                                   &            &   \checkmark                 \\ \hline
17     & Total length of forward enc TCP header                                   &            &   \checkmark                 \\ \hline
18     & Total length of backward enc TCP header                                  &            &   \checkmark                 \\ \hline
19     & No of forward enc packets                                                &            &   \checkmark                 \\ \hline
20     & No of backward enc packets                                               &            &    \checkmark                \\ \hline
21     & Total length of forward enc TCP segment                                  &            &   \checkmark                 \\ \hline
22     & Total length of backward enc TCP segment                                 &            &   \checkmark                 \\ \hline
23     & total enc payload per session                                            &            &   \checkmark                 \\ \hline
24     & IPratio enc                                                              &            &    \checkmark                \\ \hline
25-30  & Ave/Med/Max/Min/Std/Var forward enc packet length                        &            &    \checkmark                \\ \hline
31-36  & Ave/Med/Max/Min/Std/Var Interval of arrival time of forward enc traffic  &   \checkmark          &                   \\ \hline
37-42  & Ave/Med/Max/Min/Std/Var Interval of arrival time of backward enc traffic &   \checkmark          &                   \\ \hline
43-48  & Ave/Med/Max/Min/Std/Var time to live of forward enc traffic              &    \checkmark         &                   \\ \hline
49-54  & Ave/Med/Max/Min/Std/Var time to live of backward enc traffic             &    \checkmark         &                   \\ \hline
55-60  & Ave/Med/Max/Min/Std/Var TCP windows size value of forward enc traffic    &            &       \checkmark             \\ \hline
61-66  & Ave/Med/Max/Min/Std/Var TCP windows size value of backward enc traffic   &            &     \checkmark               \\ \hline
67-72  & Ave/Med/Max/Min/Std/Var length of enc IP packet                          &            &     \checkmark               \\ \hline
73-78  & Ave/Med/Max/Min/Std/Var backward enc packet length                       &            &    \checkmark                \\ \hline
79-143 & Ratio features                                                           &   -         &    -              \\ \hline
\end{tabular}
\end{table*}

\paragraph{\bf \textit{Comparison between proposed feature with other related works}}

Although Enc Feature is only applicable to encrypted traffic, it is not limited to certain encrypted protocols, and can better represent the unique characteristics of encrypted traffic itself. This makes it more suitable for encrypted traffic analysis than the protocol-specific feature. The difference among these three different feature types is summarized in Table II. Many authors have also proposed their novel feature creation approaches and grouping method to extract required features. Meghdouri et al. [11] proposed a novel cross-layer feature representation of encrypted traffic under TLS and IPSec protocols. The authors defined three modes of extracting flows: Application flows, conversation flows, and End-point flows. Zhang et al.[6] proposed a novel encoding method for protocol-specific traffic features for TLS/SSL protocols. The encoding method converts the extracted features into an image-like data format. Such an image-like feature set is fed into CNN models. Although the features they created in [11][6] are specific features for encrypted traffic, only TLS/SSL/IPSec protocols are considered. Once the traffic contains other encrypted protocols, such features are no longer applicable. Aceto et al.[28] proposed two ways to create traffic features. The first way is the 784 payload bytes of the session. The second way is to calculate the first 32 packets’ packet direction, size, TCP window size, and inter-arrival time. Bader et al.[27] proposed a novel feature creation approach which is to arrange the first 32 traffic packets into 5 groups (bidirectional packets, source, destination, handshake packets, and data transfer packets), and then based on these different group levels, more statistical features are extracted for model training. A similar feature grouping was performed by Bekerman et al.[30], the authors arrange the traffic data into 4 groups (conversation window, flow, session, transaction) to extract protocol-agnostic features. Lopez-Marti et al.[7] studied the impact of the length of traffic packets in each session flow. They proposed that keeping traffic packets between 5 to 15 packets in each session flow is the trade-off between computing time and detection performance. After ensuring that the number of traffic packets in each traffic session is consistent, then protocol-agnostic numerical feature extraction is performed. [7][27][28][30] created features all belong to protocol-agnostic numerical features which are more robust and easier to extract than protocol-specific features. However, as discussed above, traditional protocol-agnostic features only take into account the encryption properties of the session level and ignore the encryption properties of the packet level. The comparison between our proposed Enc feature and other state-of-the-art feature creation and engineering approaches is also summarized in Table III.

\begin{table}[]
\caption{The main difference among three different feature types}
\begin{tabular}{|l|l|}
\hline
\textbf{Feature Name}                                                                        & \textbf{Feature Characteristics}                                                                                                                                                           \\ \hline
\textbf{\begin{tabular}[c]{@{}l@{}}Protocol-specific \\ Features\end{tabular}}                & \begin{tabular}[c]{@{}l@{}}Features are unique to certain specific encryption \\ protocols and are only applicable to encrypted traffic.\end{tabular}                                      \\ \hline
\textbf{\begin{tabular}[c]{@{}l@{}}Traditional \\ Protocol-agnostic \\ Features\end{tabular}} & \begin{tabular}[c]{@{}l@{}}Features are not limited to certain specific encryption\\ protocols but can be extracted from both encrypted \\ traffic and non-encrypted traffic.\end{tabular} \\ \hline
\textbf{Enc Feature}                                                                         & \begin{tabular}[c]{@{}l@{}}Features are not limited to certain specific encryption\\ protocols and are only applicable to encrypted traffic.\end{tabular}                                  \\ \hline
\end{tabular}
\end{table}

\begin{table*}[]
\caption{Comparison of Feature Creation and Selection}
\centering
\begin{tabular}{|l|l|l|l|l|l|}
\hline
No. & Authors                   & \begin{tabular}[c]{@{}l@{}}Traffic Feature \\ Type\end{tabular} & \begin{tabular}[c]{@{}l@{}}Applicable \\ Protocols\end{tabular} & \begin{tabular}[c]{@{}l@{}}Consider session \\ level encryption\end{tabular} & \begin{tabular}[c]{@{}l@{}}Consider packet \\ level encryption\end{tabular} \\ \hline
1   & Zhang et al {[}6{]}       & Protocl-specific Feature                                        & TLS/SSL                                                         & YES                                                                          & YES                                                                         \\ \hline
2   & Meghdouri et al {[}11{]}  & Protocl-specific Feature                                        & TLS/IPSec                                                       & YES                                                                          & YES                                                                         \\ \hline
3   & Aceto et al {[}28{]}      & Protocol-agnostic Numerical Feature                             & ALL                                                             & YES                                                                          & NO                                                                          \\ \hline
4   & Bader et al {[}27{]}      & Protocol-agnostic Numerical Feature                             & ALL                                                             & YES                                                                          & NO                                                                          \\ \hline
5   & Bekerman et al {[}30{]}   & Protocol-agnostic Numerical Feature                             & ALL                                                             & YES                                                                          & NO                                                                          \\ \hline
6   & Lopez-Marti et al.{[}7{]} & Protocol-agnostic Numerical Feature                             & ALL                                                             & YES                                                                          & NO                                                                          \\ \hline
7   & Proposed Work                         & ENC Feature                                                     & ALL                                                             & YES                                                                          & YES                                                                         \\ \hline
\end{tabular}
\end{table*}

\section{\bf Experiment Design}
\subsection{\bf Encrypted malicious traffic detection framework}

In this section, we propose a two-layer encrypted malicious traffic detection framework comprising different deep learning and traditional machine learning algorithms. Different selected features will be trained on different appropriate algorithms to achieve effective detection results. Figure 1 is the architecture of our proposed detection framework.

\begin{figure}[h!]
\centerline{\includegraphics[width=22pc]{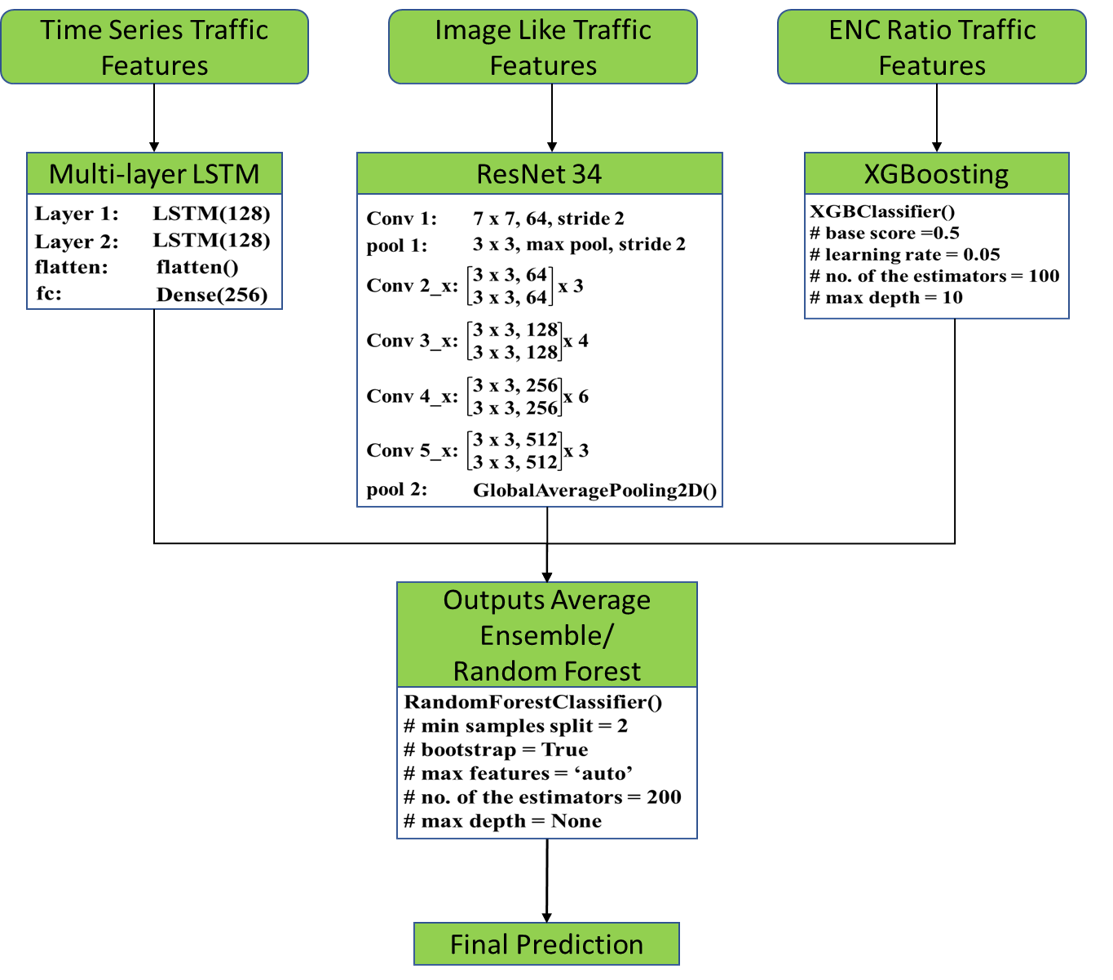}}
\caption{The Architecture of Encrypted Malicious Traffic Detection Framework with parameters.}
\end{figure}

The time-related traffic features will be selected and fed into the RNN models (GRU and LSTM). The payload-based side-channel features will be further encoded in 2D image-like format to train the CNN models (ResNet). Features related to the ratio between traditional protocol-agnostic numerical features and enc traffic features will be fed into traditional machine learning algorithms (Random Forest and XGBoost). Finally, we select the optimal performance model in each RNN, CNN, and traditional machine learning. These selected models are constructed as the layer 1 in the framework, their output probabilities will be further fed into layer 2 detector to make the final prediction. Layer 2 detector is constructed by the Random Forest method or average ensemble method depending on the importance and priority of different detection evaluation measures. The comparison and selection of optimal performance algorithms in each RNN, CNN, and traditional machine learning algorithm will be discussed in Sections IV.D, E, and F.

\subsection{\bf Dataset Collection}
In this research, to increase the variety of traffic data, we did not select only one single public dataset or generate private datasets that are not used publicly. We curated a new dataset consisting of encrypted traffic from 6 different public datasets: CTU-Malware-Capture; Benign-Capture and Mixture Capture are three datasets produced from Malware Capture Facility Project published by Stratosphere Lab [19]. CICIDS-2017 [23]; CICIDS-2012 [24]; CIRA-CIC-DoHBRW-2020 [25] are three datasets published by the Canadian Institute for Cybersecurity (CIC). 26 malicious traffic types are extracted from the CTU-Malware-Capture dataset and benign traffic is extracted from the other five datasets to balance the dataset and maximize the traffic data variety. The label of traffic (legitimate or malicious) is determined by the description of each PCAP file provided by the public dataset provider. For example, some PCAP files provided by stratosphere lab only capture the malicious traffic. The flow sessions in such PCAP files do not need to be further filtered. In contrast, some PCAP files’ description content lists infected IP addresses or the IP address of normal connections. We appropriately filter the flow sessions in such PCAP files based on the description of the PCAP file to mark them. Furthermore, as we select PCAP files from selected public datasets, we prefer to select PCAP files with a high proportion of encrypted traffic. Moreover, we also ensure that the ratio of benign and malicious traffic is roughly the same to ensure the data balance. Table IV provides detailed information about the dataset we made. To facilitate the research in this domain, Our dataset [29] is released in Mendeley Data.

\begin{table}
\caption{Details of the composed traffic dataset}
\begin{tabular}{|l|l|l|} 
\hline
\textbf{No.} & \textbf{Malicious Traffic types} & \textbf{Encrypted Session}  \\ 
\hline
1            & Ammyy                    & 14245                       \\ 
\hline
2            & Artemis Trojan           & 10246                       \\ 
\hline
3            & Barys                    & 19438                       \\ 
\hline
4            & Bunitu Botnet            & 8060                        \\ 
\hline
5            & Bunitu Botnet (Stripped) & 5560                        \\ 
\hline
6            & Caphaw/Kazy              & 24948                       \\ 
\hline
7            & Cerber Ransomware        & 26253                       \\ 
\hline
8            & Dridex                   & 6225                        \\ 
\hline
9            & HPEmotet                 & 13736                       \\ 
\hline
10           & HtBot                    & 10606                       \\ 
\hline
11           & Miuref                   & 4634                        \\ 
\hline
12           & omQUd                    & 11257                       \\ 
\hline
13           & PUA.Taobao               & 11341                       \\ 
\hline
14           & Ransom.Locky             & 26960                       \\ 
\hline
15           & Razy                     & 3207                        \\ 
\hline
16           & Sathurbot                & 1361                        \\ 
\hline
17           & TrickBot                 & 8752                        \\ 
\hline
18           & Trickster                & 11644                       \\ 
\hline
19           & Trojan.Banker            & 9296                        \\ 
\hline
20           & Trojan.Yakes             & 1820                        \\ 
\hline
21           & TrojanDownloader         & 3015                        \\ 
\hline
22           & Upatre                   & 1251                        \\ 
\hline
23           & Ursnif                   & 10552                       \\ 
\hline
24           & Vawtrak                  & 26632                       \\ 
\hline
25           & WisdomEyes               & 24228                       \\ 
\hline
26           & Zbot with others         & 11062                       \\ 
\hline
             & \textbf{Summary}         & \textbf{306329}             \\
\hline

\hline
\textbf{No.} & \textbf{Benign Datasets}                                                    & \textbf{Encrypted Session}  \\ 
\hline
1            & CIRA-CIC-DoHBRW-2020                                                        & 105524                      \\ 
\hline
2            & \begin{tabular}[c]{@{}l@{}}Benign Capture and\\Mixture Capture\end{tabular} & 79619                       \\ 
\hline
3            & CICIDS-2017                                                                 & 92975                       \\ 
\hline
4            & CICIDS-2012                                                                 & 26209                       \\ 
\hline
             & \textbf{Summary}                                                            & \textbf{304327}             \\
\hline
\end{tabular}

\end{table}

\subsection{\bf Experiment Setup and Performance Evaluation}
The experiment running: Intel(R) Core(TM) i7-10700K CPU @ 3.8GHz 64.0GB of RAM. In order to construct our models, scikit-learn, TensorFlow, and XGBoost libraries for Python are used. When evaluating the performance of detection models, accuracy, F1, precision, recall, AUC-ROC, False Positive Rate (FPR), and True Positive Rate (TPR) are selected to provide a more comprehensive analysis of performance results. They are deﬁned as in the following equations:

\[Accuracy =\frac{TP+TN}{Total}\]
\[True Positive Rate/ Recall =\frac{TP}{TP+FN}\]
\[False Positive Rate =\frac{FP}{FP+TN}\]
\[Precision = \frac{TP}{TP+FP}\]
\[F1score =2 *\frac{precision * recall}{precision + recall}\]

The detection of encrypted traffic is treated as a dichotomy. Encrypted malicious traffic is viewed as the positive sample to be classified, while benign traffic is viewed as the negative sample. Therefore, True positive(TP) is the number of encrypted malicious traffic classified as malicious. True negative(TN) is the number of benign encrypted traffic classified as benign. False positive(FP) is the number of benign encrypted traffic classified as malicious. False negative(FN) is the number of encrypted malicious traffic identiﬁed as benign.

\subsection{\bf Feature Extraction and Data Pre-processing}
Firstly, the dataset will be filtered through Wire-Shark, the purpose of this process is to filter out the unencrypted traffic. The filtered PCAP files will be fed into our feature extraction function to extract the features we need. The feature extraction function contains more than 300 different feature extraction logic. The function can extract both session-based and packet-based traffic features with high efficiency. Such features include time-related features (i.e., flow duration of forward traffic and Interval of the arrival time of backward traffic), payload side-channel features (i.e, Payload ratio, length of TCP payload, and total payload per session), our proposed enc features (i.e., total time to live of forward encrypted traffic in each encrypted session and total length of encrypted IP packet in each encrypted session), and enc ratio features through calculating the ratio between the traditional protocol-agnostic feature and enc feature (i.e, Ratio of flow duration between encrypted and non-encrypted traffic in each encrypted session). A more detailed dataset collection and extraction process can be found in our publicly available datasets [29] and the flow chart of the feature creation function is shown in Fig 9 of the Appendix.

The next step is the data pre-processing. Our data pre-processing can be divided into three steps. The first step is the packet number cut off in each session. In our dataset, all sessions with more than 15 packets will only keep the first 15 packets. Previous research [7] has considered the impact of the length of traffic session flow and the trade-off between computing time and detection rate. Their experiment has indicated that keeping to between five and fifteen packets in each session can already achieve a relatively good performance.

The second step is to do the simulation padding. For those sessions without enough 15 packets will be padded based on average padding. The calculation method is to calculate the average value of all packets in the same session for each feature, and then pad this average value into the session until the packet number reaches 15 packets. 

The third step is to design the final data input format according to the characteristics of different detection models, which we will discuss in the following parts respectively.

\paragraph{\bf \textit{time-related Features with RNN models}}

Based on the discussion in the section above, in this experiment, we will choose the appropriate model according to the characteristics of the features, instead of choosing the model first and then adjusting the features. We first selected 85 time-related features. A 15x85 (85 features of 15 chronologically arranged packets with 85 features in each flow session) time-related input format data, Figure 2, will be generated. These features will be fed into LSTM and GRU models for training.

\begin{figure}
\centerline{\includegraphics[width=14pc]{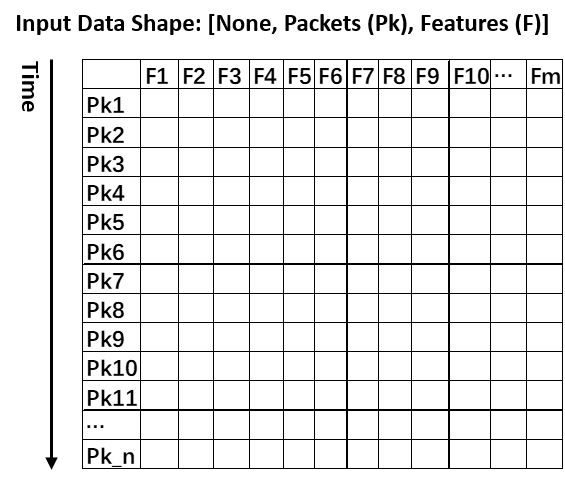}}
\caption{The Structure of time-related input format data. (m = 85 which means 85 time-related features are selected; n = 15 means to keep the first 15 packets of each session in chronological order. The intersection is the value of the $m$th feature in the $n$th packet).}
\end{figure}

\textbf{LSTM} stands for long short-term memory networks. It is a special type of RNN, specifically designed to overcome the long-term dependency problem faced by the recurrent neural network (RNN). It can handle the gradient vanishing problem faced by RNN. The core concept of LSTM is the cell state and the "gate" structure. The cell states are equivalent to the paths of information transmission, allowing information to be passed on in a sequence. Theoretically, the cell state is able to pass on the relevant information from the sequence processing all the way through. Thus, even information from earlier time steps can be carried to cells of later time steps, which overcomes the effects of short-term memory. The "gate" structure (forget gate, input gate, and output gate) learns which information to keep or forget during the training process. Therefore, LSTM is particularly good at processing text, speech, and time-related analysis. Bidirectional LSTM means that each training sequence is presented both forward and backward. Both sequences are connected to the same output layer. The two-way LSTM has complete information about each point in a given sequence, as well as information before and after it.

\textbf{GRU} stands for Gated Recurrent Unit. Both GRU and LSTM are designed to solve the gradient vanishing problem that occurs in standard recurrent neural networks. They both have internal mechanisms called gates that regulate the flow of information. GRU is a newer generation of recurrent neural networks and is similar to LSTM. But GRU removes the cell states and uses hidden states to transmit information. It also has only two gates (reset gate and update gate). GRU can also be considered as a variant of LSTM.

\begin{figure}
\centerline{\includegraphics[width=22pc]{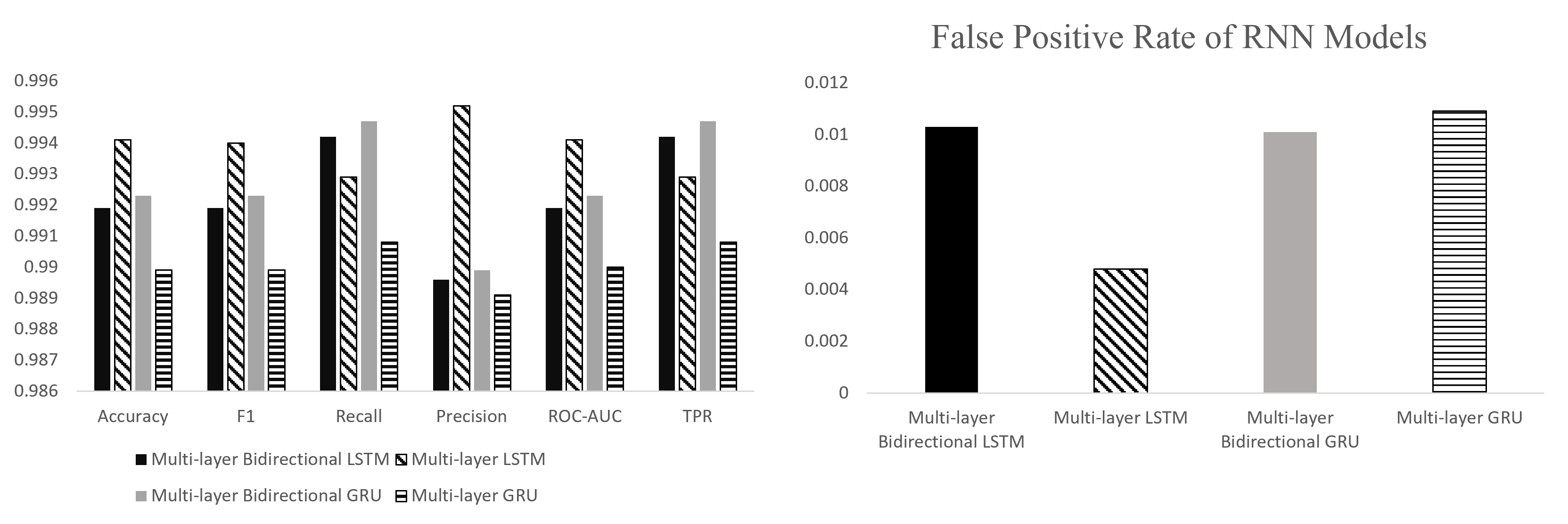}}
\caption{The experimental results of time-related traffic features with RNN models}
\end{figure}

The experimental results of RNN models are plotted in Figure 3 and exact values are shown in Appendix Table VI. According to the experimental results, the multi-layer LSTM and multi-layer bidirectional GRU outperform the other two models (multi-layer bidirectional LSTM and multi-layer GRU). Although multi-layer bidirectional GRU achieved better Recall/TPR (0.18\% higher than multi-layer LSTM), its accuracy, F1 score, precision, and FPR are all worse than multi-layer LSTM. In particular, the multi-layer LSTM’s precision score is 0.53\% higher and FPR is 0.53\% lower than that of the multi-layer bidirectional GRU. Therefore, the multi-layer LSTM model is selected as the model for training time-related traffic input data in the detection framework.

\paragraph{\bf \textit{Payload based side-channel image like data input with CNN models}}

When each flow session is expanded according to the first 15 packets, a 2D image-like array (with [15 X Feature number]), Figure 4, can be obtained, which can be fed into CNN models as a single-channel image. In the experiment, we selected 38 payload-based side-channel features to generate a 15 x 38 image-like array input. Simultaneously, we also proposed a 2D image generation method, which is to use matrix multiplication (38 x 15 @ 15 x 38 = 38 x 38) to generate a square 2D array, Figure 5, with equal length and width. It can be fed into CNN models as another single-channel image format. Both types of input data formats will be fed into ResNet models and the experimental results will be compared.

\begin{figure}
\centerline{\includegraphics[width=21pc]{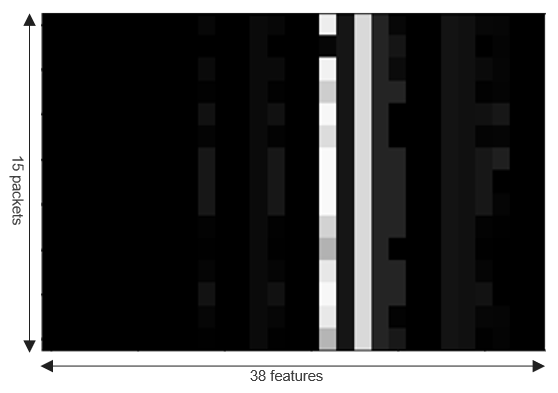}}
\caption{The Structure of 15 x 38 image-like array input}
\end{figure}

\begin{figure}
\centerline{\includegraphics[width=15pc]{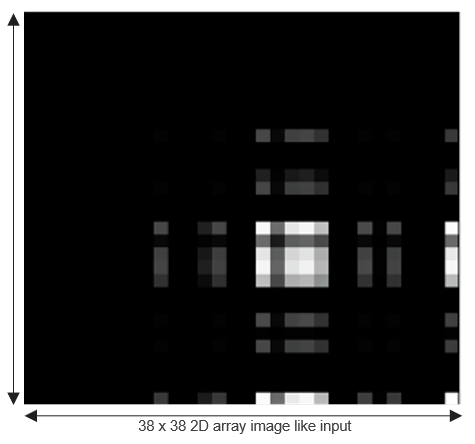}}
\caption{The Structure of square 2D array input (38 x 15 @ 15 x 38 = 38 x 38)}
\end{figure}

\textbf{ResNet} model stands for residual neural network, which is a special type of neural network that was first introduced by He et al.[26] in 2015. The model applies a concept called residual blocks and a technique called skip connection to solve the problem of gradient vanishing or exploding. A residual block is constructed by skip connections that connect the activation of one layer to further layers by skipping some of the intermediate layers. Thus, if any layer impairs the performance of the model then such layers will be skipped by regularisation. ResNet model is built by stacking such residual blocks so as to achieve a very deep feed-forward neural network model which can contain hundreds of layers.

\begin{figure}[h!]
\centerline{\includegraphics[width=24pc]{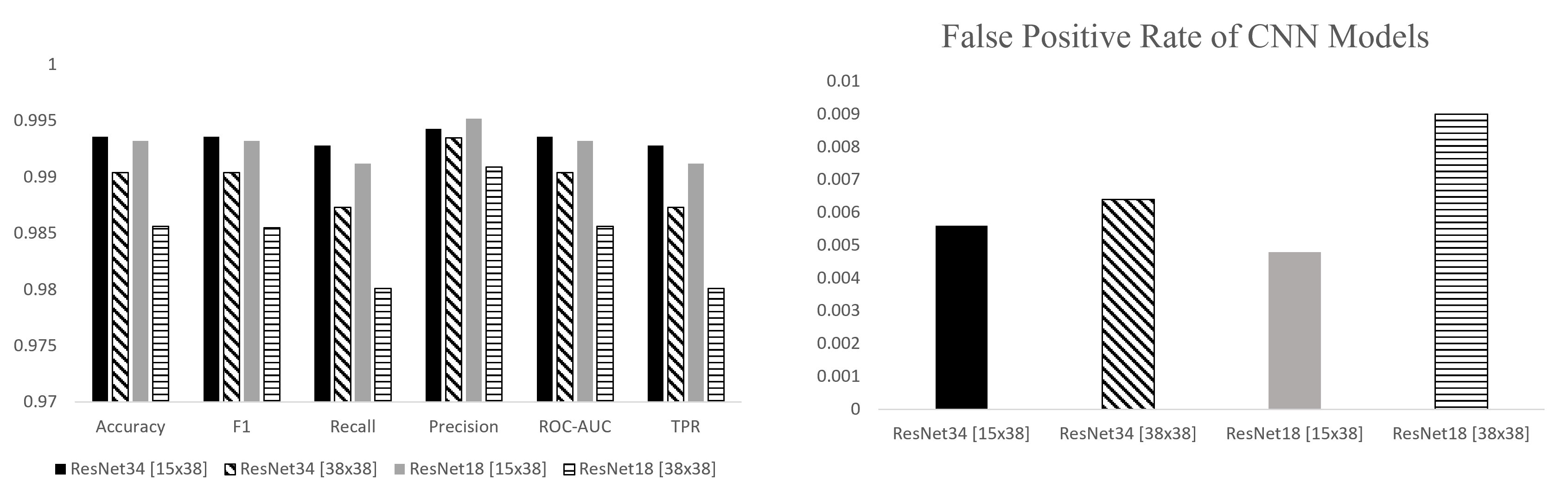}}
\caption{The experimental results of payload-based side channel traffic features with ResNet models}
\end{figure}

The experimental results are plotted in Figure 6 and exact values are shown in Appendix Table VII. According to the experimental results, ResNet 34 and ResNet 18 models with 15x38 data inputs can achieve the top 2 detection performance compared with other ResNet models in terms of all performance evaluations. ResNet 18 achieved the highest 99.52\% precision value and 0.48\% FPR. ResNet 34 achieved the highest performance in terms of accuracy, f1, recall, ROC-AUC, and TPR among all ResNet models. Therefore, we choose the ResNet34 model (with 15x38 data input format) as the final model for training with payload-based side-channel traffic feature in the detection framework.

\paragraph{\bf \textit{Ratio Features between Enc Features and Traditional protocol-agnostic numerical features with traditional machine learning models}}

This set of features is the ratio of feature values calculated according to the relationship between enc feature and its corresponding traditional protocol-agnostic numerical feature. A total of 74 ratio features were selected to form the ratio feature set. These ratio features will be fed into Random Forest and XGBoost algorithms for training, and the detection performance will be compared. 

\textbf{Random Forest} is a supervised machine learning algorithm, which uses both the bagging method and feature randomness to generate a forest of uncorrelated decision trees. Random Forest applies majority vote in classification and mean vote in regression to obtain a more accurate and reliable prediction. The random forest algorithm has three main hyper-parameters that need to be set before training. These parameters include n\_estimators, max\_features, and min\_sample\_leaf.

\textbf{XGBoost} stands for eXtreme Gradient Boosting. XGBoost is an implementation of gradient boosting decision trees designed to improve speed and performance. In this algorithm, decision trees are created in a sequential form. All independent variables are assigned weights and then fed into decision trees for training to make predictions. These individual classifiers or predictors are then aggregated to give a more accurate result.

\begin{figure}[h!]
\centerline{\includegraphics[width=22pc]{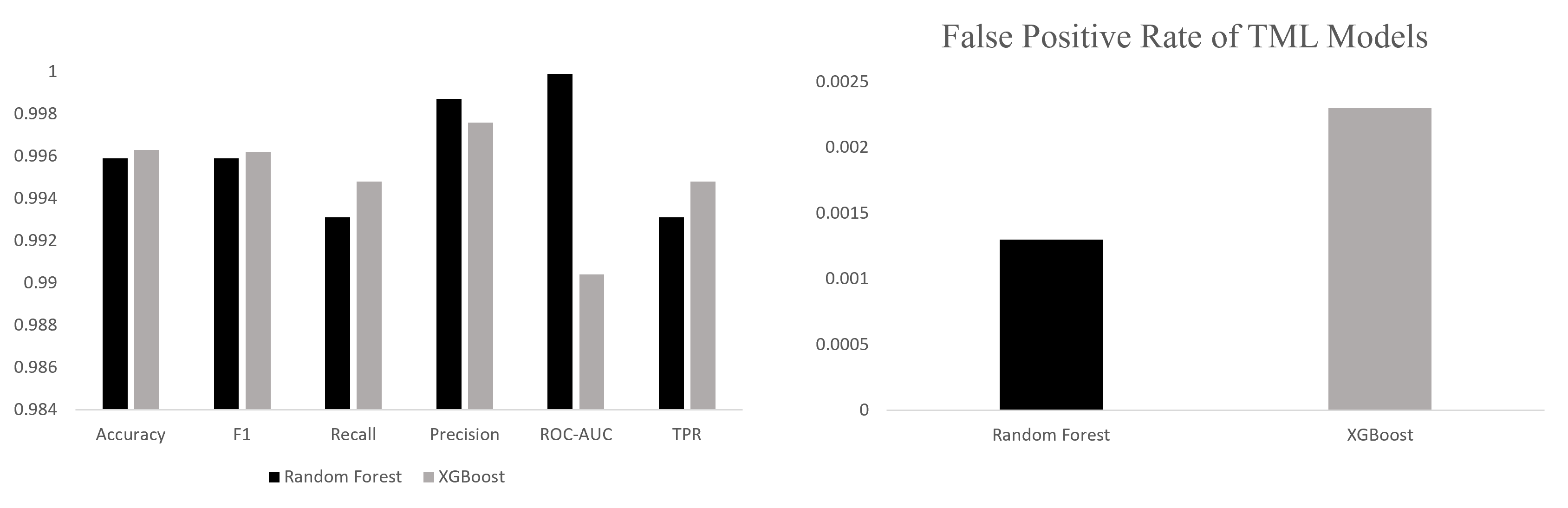}}
\caption{The experimental results of enc ratio traffic features with Traditional Machine Learning (TML) models}
\end{figure}

The reason we chose Random Forest and XGBoost is that both models are among the top-performing algorithms in traditional machine learning in many previous studies. The experimental results are plotted in Figure 7 and exact values are shown in Appendix Table VIII. Both Random Forest and XGboost achieved above 99\% performance rates in terms of accuracy, F1, precision, recall, ROC-AUC, and TPR. Their FPR are both below 1\%. According to the comparison of the experimental results, the XGBoost model performs slightly better than Random Forest in terms of accuracy, f1, recall, ROC-AUC, and TPR. In contrast, random forest is only precision 0.11\% higher and FPR 0.10\% lower than XGBoost. Therefore, we choose the XGBoost model as the final traditional machine learning model for using ratio feature input data in the detection framework.

\subsection{\bf Experiment Result and Evaluation}
The algorithms used to train the three different feature types in the detection framework have been selected through the above comparative experiments. They will output their respective prediction probabilities according to the respective input feature set, and then these predictions will be fed into the layer 2 final detector of the detection framework. The Random Forest method or average ensemble method is used as the layer 2 final detector in the proposed detection framework. In practice, the appropriate layer 2 detector method can be chosen according to the importance and priority of different evaluation measures. The final detection results of the encrypted malicious traffic detection framework are plotted in Figure 8 and exact values are shown in Appendix Table IX. According to the experiment results, while we select random forest as the final layer 2 detector in the framework, the framework can achieve the highest 99.68\% TPR value which outperforms above deep learning and traditional machine learning algorithms. On the other hand, as we use the average ensemble method as the layer 2 detector in the framework, it can achieve the highest 99.73\% accuracy, 99.72\% F1 score, and 99.89\% precision. It also achieves the lowest 0.11\% FPR in the experiment.

\begin{figure}[h!]
\centerline{\includegraphics[width=22pc]{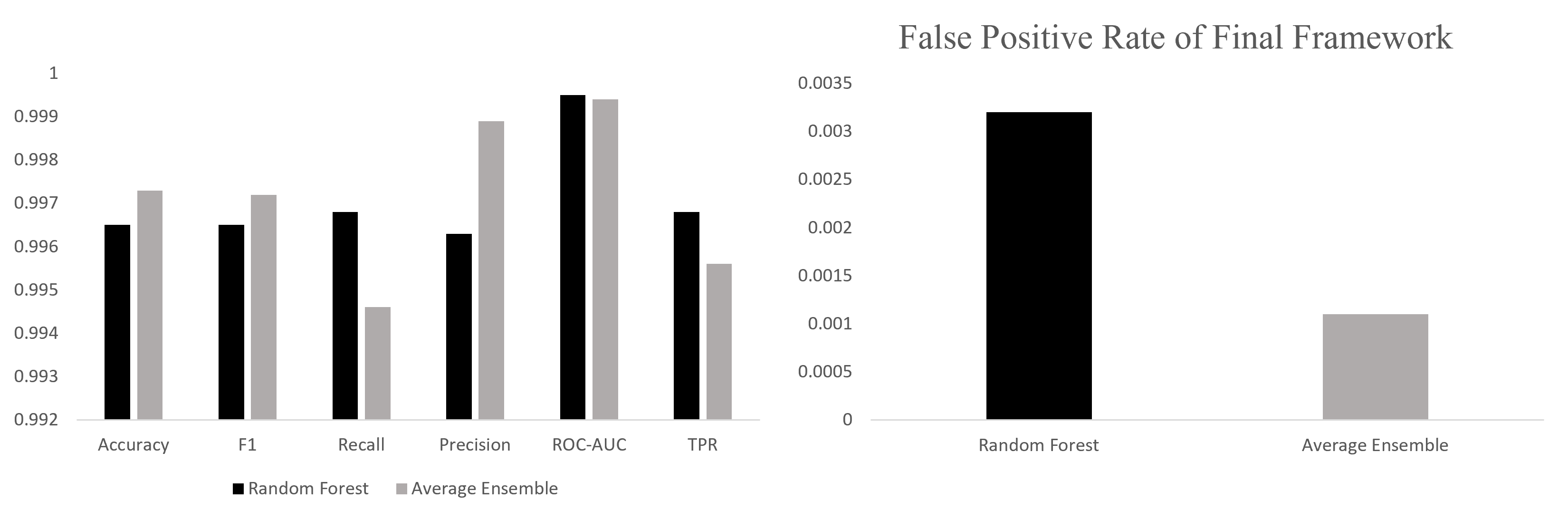}}
\caption{The final experimental results of the encrypted malicious traffic detection framework}
\end{figure}

\subsection{\bf Comparative Experiments With or Without Proposed Enc Features}
In Section III, we analyze the advantages of ENC features compared with traditional traffic features. For example, Enc features are more representative of the data distribution characteristics of encrypted traffic than other traditional selected features and are no longer limited by the type of encryption protocol. It also provides more selectable features for the feature set used in machine learning algorithms. In this subsection, we conduct a series of comparative experiments to verify the performance improvement of Enc features in the proposed machine learning based encrypted traffic detection framework. Two different feature sets are used to train our detection framework to conduct comparative experiments, one is the feature set used in the above experiments, which contains both traditional protocol-agnostic features and proposed Enc features. The other feature set is to remove the Enc features and then feed the remaining features into the detection framework. The results of comparative experiments are recorded in Table V. According to the experimental results, we can find that after the removal of Enc features, no matter whether we choose the Random Forest method or average ensemble method as the layer 2 final detector in the detection framework, the experimental results are worse than the detection framework with Enc features. Feature set contains enc features in the detection framework with Random Forest as layer 2 detector achieved highest 99.68\% Recall value. Feature set contains enc features in the detection framework with average ensemble method as layer 2 detector achieved highest 99.73\% accuracy, 99.72\% F1 score, and lowest 0.11\% false positive rate.

\begin{table*}[h!]
\centering
\caption{The Comparative Experiment Results of The Detection Framework With or Without Proposed Enc Features.}
\begin{tabular}{llllllll}
\hline
\multicolumn{1}{|l|}{\textbf{Selected layer 2 detector in detection framework}} & \multicolumn{1}{l|}{Accuracy}        & \multicolumn{1}{l|}{F1}              & \multicolumn{1}{l|}{Recall}          & \multicolumn{1}{l|}{Precision}       & \multicolumn{1}{l|}{ROC-AUC}         & \multicolumn{1}{l|}{FPR}             & \multicolumn{1}{l|}{TPR}             \\ \hline
\multicolumn{1}{|l|}{Random Forest detector option with enc features}                      & \multicolumn{1}{l|}{99.65}          & \multicolumn{1}{l|}{99.65}          & \multicolumn{1}{l|}{\textbf{99.68}} & \multicolumn{1}{l|}{99.63}          & \multicolumn{1}{l|}{\textbf{99.95}} & \multicolumn{1}{l|}{0.32}          & \multicolumn{1}{l|}{\textbf{99.68}} \\ \hline
\multicolumn{1}{|l|}{Random Forest detector option without enc features}                   & \multicolumn{1}{l|}{99.33} & \multicolumn{1}{l|}{99.33} & \multicolumn{1}{l|}{99.32}          & \multicolumn{1}{l|}{99.34} & \multicolumn{1}{l|}{99.84}          & \multicolumn{1}{l|}{0.65} & \multicolumn{1}{l|}{99.32}          \\ \hline   
\multicolumn{1}{|l|}{Average Ensemble detector option with enc features}                   & \multicolumn{1}{l|}{\textbf{99.73}} & \multicolumn{1}{l|}{\textbf{99.72}} & \multicolumn{1}{l|}{99.46}          & \multicolumn{1}{l|}{\textbf{99.89}} & \multicolumn{1}{l|}{99.94}          & \multicolumn{1}{l|}{\textbf{0.11}} & \multicolumn{1}{l|}{99.56}          \\ \hline  
\multicolumn{1}{|l|}{Average Ensemble detector option without enc features}                   & \multicolumn{1}{l|}{99.45} & \multicolumn{1}{l|}{99.44} & \multicolumn{1}{l|}{99.25}          & \multicolumn{1}{l|}{99.64} & \multicolumn{1}{l|}{99.45}          & \multicolumn{1}{l|}{0.36} & \multicolumn{1}{l|}{99.25}          \\ \hline  
\end{tabular}
\end{table*}

\section{\bf Conclusion}

While encryption mechanisms protect the privacy of users, adversaries also apply encryption mechanisms to hide their malicious intent. The contribution of this paper is to conduct a comprehensive analysis of traffic characteristics of encrypted traffic and different traffic feature creation approaches. A new feature creation approach is also proposed that can be more appropriate to represent the distribution characteristics of encrypted traffic and generate an encrypted traffic feature list that contains 143 Enc features. Finally, different sets of features are fed into our proposed encrypted malicious traffic detection framework which is constructed by both deep learning and traditional machine learning model. The experimental results demonstrate that the performance of our framework outperforms any kind of classical deep learning or traditional machine learning models used alone in the framework such as ResNet, LSTM, and Random Forest detection models. The proposed detection framework achieved a 99.72\% F1 score, 0.11\% FPR, and 99.56\% TPR. In the future, we will further study encrypted malicious traffic samples and their representative features to further improve the effectiveness and performance of encrypted malicious traffic detection and multi-class classification. More different machine learning and deep learning algorithms will be also designed into the detection framework.

\onecolumn
\section*{Appendix}

\begin{table*}[h!]
\centering
\caption{The experimental results of time-related traffic features with RNN models}
\begin{tabular}{|l|l|l|l|l|l|l|l|}
\hline
\textbf{time-related Features}  & Accuracy        & F1              & Recall          & Precision       & ROC-AUC         & FPR             & TPR             \\ \hline
Multi-layer Bidirectional LSTM & 99.19          & 99.19          & 99.42          & 98.96          & 99.19          & 1.03          & 99.42          \\ \hline
Multi-layer LSTM               & \textbf{99.41} & \textbf{99.40} & 99.29          & \textbf{99.52} & \textbf{99.41} & \textbf{0.48} & 99.29          \\ \hline
Multi-layer Bidirectional GRU  & 99.23          & 99.23          & \textbf{99.47} & 98.99          & 99.23          & 1.01          & \textbf{99.47} \\ \hline
Multi-layer GRU                & 98.99          & 98.99          & 99.08          & 98.91          & 99.00          & 1.09          & 99.08          \\ \hline
\end{tabular}
\end{table*}

\begin{table*}[h!]
\centering
\caption{The experimental results of payload-based side channel traffic features with ResNet models}
\begin{tabular}{llllllll}
\hline
\multicolumn{1}{|l|}{\textbf{Image-like Features}} & \multicolumn{1}{l|}{Accuracy}        & \multicolumn{1}{l|}{F1}              & \multicolumn{1}{l|}{Recall}          & \multicolumn{1}{l|}{Precision}       & \multicolumn{1}{l|}{ROC-AUC}         & \multicolumn{1}{l|}{FPR}             & \multicolumn{1}{l|}{TPR}             \\ \hline
\multicolumn{1}{|l|}{ResNet34 {[}15x38{]}}         & \multicolumn{1}{l|}{\textbf{99.36}} & \multicolumn{1}{l|}{\textbf{99.36}} & \multicolumn{1}{l|}{\textbf{99.28}} & \multicolumn{1}{l|}{99.43}          & \multicolumn{1}{l|}{\textbf{99.36}} & \multicolumn{1}{l|}{0.56}          & \multicolumn{1}{l|}{\textbf{99.28}} \\ \hline
\multicolumn{1}{|l|}{ResNet34 {[}38x38{]}}         & \multicolumn{1}{l|}{99.04}          & \multicolumn{1}{l|}{99.04}          & \multicolumn{1}{l|}{98.73}          & \multicolumn{1}{l|}{99.35}          & \multicolumn{1}{l|}{99.04}          & \multicolumn{1}{l|}{0.64}          & \multicolumn{1}{l|}{98.73}          \\ \hline
\multicolumn{1}{|l|}{ResNet18 {[}15x38{]}}         & \multicolumn{1}{l|}{99.32}          & \multicolumn{1}{l|}{99.32}          & \multicolumn{1}{l|}{99.12}          & \multicolumn{1}{l|}{\textbf{99.52}} & \multicolumn{1}{l|}{99.32}          & \multicolumn{1}{l|}{\textbf{0.48}} & \multicolumn{1}{l|}{99.12}          \\ \hline
\multicolumn{1}{|l|}{ResNet18 {[}38x38{]}}         & \multicolumn{1}{l|}{98.56}          & \multicolumn{1}{l|}{98.55}          & \multicolumn{1}{l|}{98.01}          & \multicolumn{1}{l|}{99.09}          & \multicolumn{1}{l|}{98.56}          & \multicolumn{1}{l|}{0.90}          & \multicolumn{1}{l|}{98.01} \\ \hline                                 
\end{tabular}
\end{table*}

\begin{table*}[h!]
\centering
\caption{The experimental results of enc ratio traffic features with RF and XGBoost models}
\begin{tabular}{llllllll}
\hline
\multicolumn{1}{|l|}{\textbf{Enc\_ratio Features}} & \multicolumn{1}{l|}{Accuracy}        & \multicolumn{1}{l|}{F1}              & \multicolumn{1}{l|}{Recall}          & \multicolumn{1}{l|}{Precision}       & \multicolumn{1}{l|}{ROC-AUC}         & \multicolumn{1}{l|}{FPR}             & \multicolumn{1}{l|}{TPR}             \\ \hline
\multicolumn{1}{|l|}{Random Forest}                & \multicolumn{1}{l|}{99.59}          & \multicolumn{1}{l|}{99.59}          & \multicolumn{1}{l|}{99.31}          & \multicolumn{1}{l|}{\textbf{99.87}} & \multicolumn{1}{l|}{99.989}         & \multicolumn{1}{l|}{\textbf{0.13}} & \multicolumn{1}{l|}{99.31}          \\ \hline
\multicolumn{1}{|l|}{XGBoost}                      & \multicolumn{1}{l|}{\textbf{99.63}} & \multicolumn{1}{l|}{\textbf{99.62}} & \multicolumn{1}{l|}{\textbf{99.48}} & \multicolumn{1}{l|}{99.76}          & \multicolumn{1}{l|}{\textbf{99.99}} & \multicolumn{1}{l|}{0.23}          & \multicolumn{1}{l|}{\textbf{99.48}} \\ \hline                                  
\end{tabular}
\end{table*}

\begin{table*}[h!]
\centering
\caption{The final experimental results of the encrypted malicious traffic detection framework}
\begin{tabular}{llllllll}
\hline
\multicolumn{1}{|l|}{\textbf{Selected layer 2 detector in the framework}} & \multicolumn{1}{l|}{Accuracy}        & \multicolumn{1}{l|}{F1}              & \multicolumn{1}{l|}{Recall}          & \multicolumn{1}{l|}{Precision}       & \multicolumn{1}{l|}{ROC-AUC}         & \multicolumn{1}{l|}{FPR}             & \multicolumn{1}{l|}{TPR}             \\ \hline
\multicolumn{1}{|l|}{Random Forest}                      & \multicolumn{1}{l|}{99.65}          & \multicolumn{1}{l|}{99.65}          & \multicolumn{1}{l|}{\textbf{99.68}} & \multicolumn{1}{l|}{99.63}          & \multicolumn{1}{l|}{\textbf{99.95}} & \multicolumn{1}{l|}{0.32}          & \multicolumn{1}{l|}{\textbf{99.68}} \\ \hline
\multicolumn{1}{|l|}{Average Ensemble}                   & \multicolumn{1}{l|}{\textbf{99.73}} & \multicolumn{1}{l|}{\textbf{99.72}} & \multicolumn{1}{l|}{99.46}          & \multicolumn{1}{l|}{\textbf{99.89}} & \multicolumn{1}{l|}{99.94}          & \multicolumn{1}{l|}{\textbf{0.11}} & \multicolumn{1}{l|}{99.56}          \\ \hline                               
\end{tabular}
\end{table*}

\begin{figure}
\centerline{\includegraphics[width=42pc]{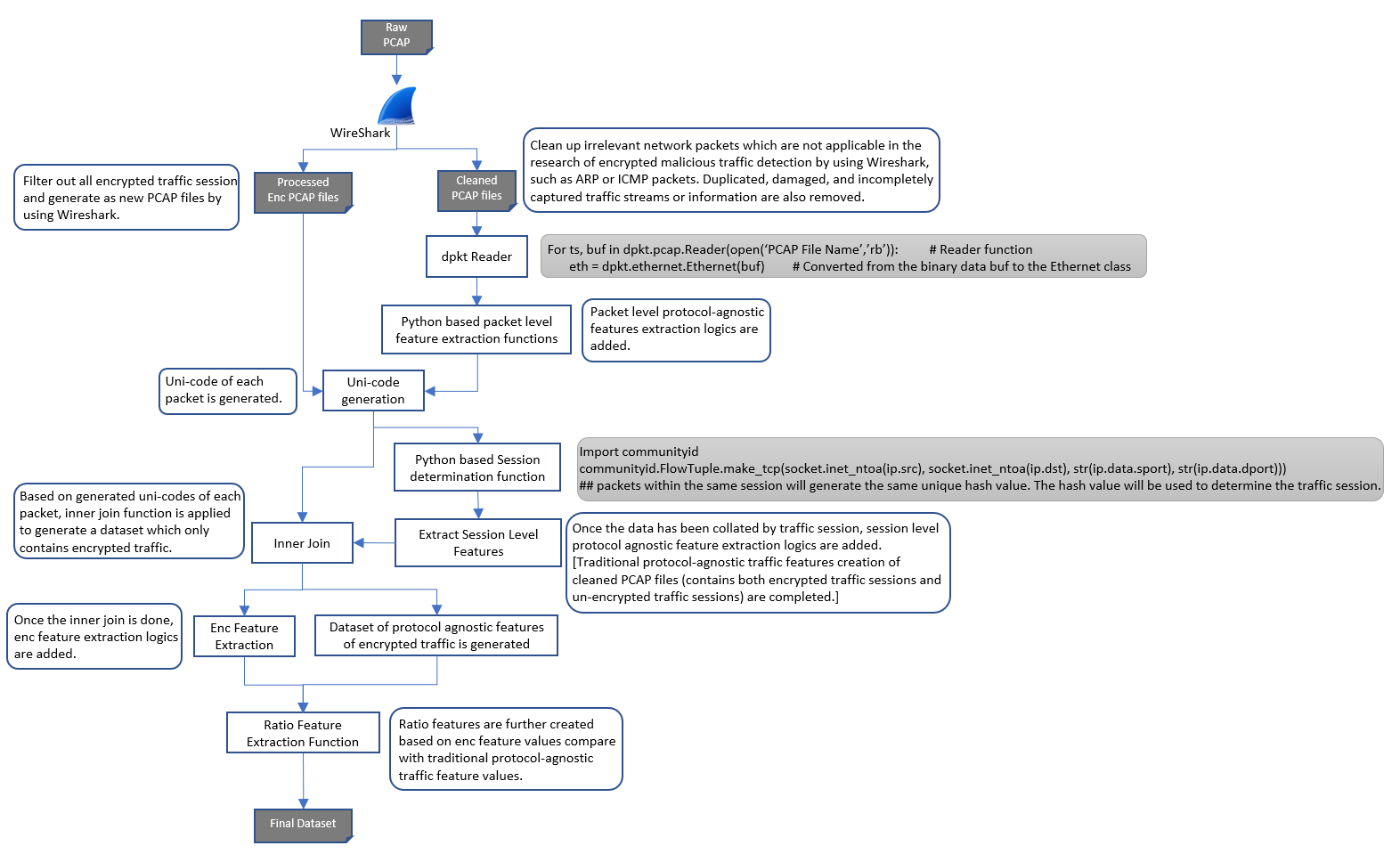}}
\caption{Flow Chart of the proposed feature creation function}
\end{figure}

\end{document}